**Atomic bonding and electrical potential at metal/oxide interfaces, a first principle study**

Eric Tea[1], Jianqiu Huang[1], Guanchen Li[1], and Celine Hin[1,2]

[1] Department of Mechanical Engineering, Virginia Tech, Goodwin Hall, 635 Prices Fork Road - MC 0238, Blacksburg, VA 24061, USA

[2] Department of Material Science and Engineering, Virginia Tech, Goodwin Hall, 635 Prices Fork Road - MC 0238, Blacksburg, VA 24061, USA

**ABSTRACT**

A number of electronic devices involve metal/oxide interfaces in their structure where the oxide layer plays the role of electrical insulator. As the downscaling of devices continues, the oxide thickness can spread over only a few atomic layers, making the role of interfaces prominent on its insulating properties. The prototypical $Al/SiO_2$ metal/oxide interface is investigated using first principle calculations, and the effect of the interfacial atomic bonding is evidenced. It is shown that the interface bonding configuration critically dictates the mechanical and electronic properties of the interface. Oxygen atoms are found to better delimit the oxide boundaries than cations. Interfacial cation-metal bonds allow the metal potential to leak inside the oxide layer, without atomic diffusion, leading to a virtual oxide thinning.

**INTRODUCTION**

Metal/oxide interfaces are present in many electronic devices such as Field Effect Transistors (FETs), Resistive Random Access Memory devices (ReRAM) or Tunnel Junctions (TJs). The downscaling of FETs has brought many reliability challenges. In particular, the reduced dimensions of gate oxides push the insulating materials to their classical limit, where quantum effects such as electron tunneling emerge. This causes the gate current leakage to increase, which can further be enhanced by the presence of point defects that provide shallow traps for tunneling, ultimately triggering oxide dielectric breakdown[1-3]. While being a reliability issue in transistors, the dielectric breakdown is vital to ReRAM device operation[4-6]. The recent interest in these memory devices has brought a significant focus on metal-oxide materials, where the diffusion of atoms during the electroforming phase is crucial[7,8]. In TJs, electron tunneling processes are desired and oxide thicknesses are kept low for that purpose. The very low oxide thickness makes the oxide layer very sensitive to interface effects such as interface charging[9], or potential barrier modification caused by Metal Induced Gap States (MIGS)[10]. In each case, the electrical degradation process of the oxide layer has to be deeply understood to be carefully controlled.

A variety of models have been proposed to explain the time dependence of dielectric breakdown observed in transistors. The knowledge of the time to breakdown is used to extrapolate the lifetime of electronic devices. All these models correlate the time dependence of the breakdown to the dynamics of point defect generation in thin oxides[1-3]. Both potential-based and current-based breakdown models assume that the applied electric field, or the flow of tunneling and thermally injected electron in the oxide barrier, is responsible for the creation of point defects[11,12]. It is further assumed that such defects act as charged traps that locally enhance the oxide electric field. In the intrinsic time dependent breakdown of silica, the involved point defects have been identified as dangling bonds[2,13]. The generation of these point defects is facilitated by the application of an electric field, that stretches the polar Si-O bonds until breakage, ultimately accelerating the time to breakdown. While being successful at reproducing the time to breakdown trends as a function of the applied electric magnitude, adjusted parameters are most often used to describe the bond breaking dynamics. In experiments, impurities present in the oxide or diffusing from the electrodes can trigger extrinsic breakdown, and the precise nature of the involved point defects is rarely assessed[11,12]. The point defect characteristics can however be deduced from measurements, where it has been concluded that involved defects in SiCOH low-k dielectrics must be positively charged and able to trap electrons. In this case, the possible dangling bonds can result from Si-O or Si-H bond breaking. Carbon dangling bonds are also possible candidates.

Time dependent dielectric breakdown also varies with the oxide thickness, which has been explained using the percolation path model[14,15]. In this model, it is assumed that the random generation of point defects creates a network of electron traps. When the trap density is high enough, a conduction path is formed, triggering the dielectric breakdown. This model successfully reproduced the decrease in critical defect density observed at breakdown when the oxide thickness is decreased.

In ReRAM devices, the necessary soft breakdown is caused by the formation of a conductive filament composed by oxygen vacancies, or by interstitial metal atoms diffusing from the electrodes, depending on the material used. In the oxide layer, both oxygen vacancies and metal atoms can be viewed as charged species. The diffusion of these defects, and possible associated bond breakage, is driven by the applied electric field. Recent studies showed that the shape of the conductive filament obtained after the electroforming stage strongly impacts the set and reset voltages[5-8,16]. This stresses the importance of local atomic environment on point defect diffusion for an accurate understanding of device stability.

The breakdown observed in the different device families rely on the creation and diffusion of point defects in the oxide layer, that is driven by the applied electric field. In ultra-thin oxides, the insulating layer thickness can be brought down to 1 or 2 nm, only a few atomic layers. In such conditions, the width of the electrode/oxide interfacial atomic layers can constitute up to 10% of the total oxide thickness. Hence, the morphology of the interfaces is critical to an accurate determination of the electric field in the oxide. Moreover, bond breakage models developed for interface regions are then necessary for an accurate description of breakdown. The available electric field dependent polar bond breakage models developed and tailored for bulk $SiO_2$ do not account for the change in nearest neighbors at metal/oxide interfaces.

In the following, the effect of the morphology of metal/oxide interfaces on the mechanical and electronic properties is explored using Density Functional Theory calculations. The study is applied to the prototypical $Al/SiO_2$ interface, representative of metal/silica-based oxide junctions. The studied system is described in the methodology section. The link between atomic bond formation at the interface, atomic charge transfer, and interface adhesion is established in the result section. The effect of interface morphology, that is the atomic local environment, is studied by the introduction of interfacial oxygen vacancies. While an oxygen vacancy in a crystalline solid is a point defect, its generation dynamic has not been studied in the present work, and the purpose of the introduced interfacial oxygen vacancies is to reveal cation-metal bonds. The response to external electric fields is monitored to depict the interface electronic properties. The definition of the interface position is also questioned, in view of the atom positions and the electrostatic potential. The discussion section set our findings in the light of the current understanding of dielectric breakdown in various systems.

**METHODS**

The slab geometry has been adopted to study an isolated $Al/SiO_2$ interface. Figure 1a depicts the supercell with the interface between the oxygen-terminated $SiO_2$ (001) slab, and Al (111) slab. The interface is characterized by a strain below 1%, allowing for a clear picture of the interface properties. Periodic slab images are separated by a 20 Å vacuum layer to minimize the surface slab-slab interaction. The open surfaces are passivated with hydrogen atoms on the $SiO_2$ side, and by 0.75 fractional hydrogen atoms on the Al side, in order to lock the charge density close to the surfaces in a bulk-like environment. To study the effect of local atomic environment at the interface, oxygen vacancies are introduced by removing one oxygen atom at the interface from the supercell. Supercells with a 83.6 Å$^2$ cross section have been used, corresponding to a $1.2\times10^{14}$ cm$^{-2}$ oxygen vacancy areal concentration that is representative of defective crystalline interfaces[17]. The modelled interface is characterized by the following vertical stacking sequence: O – Si – O – Al – Al – Al. These 6 atomic layers are viewed from the $SiO_2$ side of the interface in Figure 1b, showing the four interfacial subunits per supercell. Each pattern is composed of one interfacial Si atom bonded to three oxygen atoms, one of which being an interfacial oxygen attached to two aluminum atoms. Each pattern has three interfacial aluminum atoms.

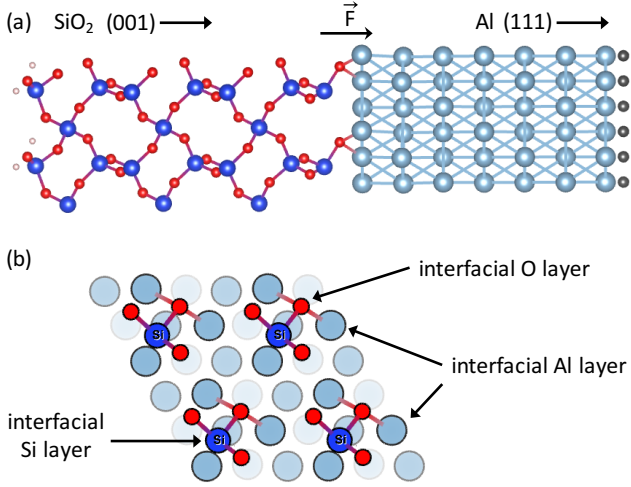

Figure 1: (a) Side view of the $SiO_2$/Al slab showing the interface and passivated surfaces. (b) Top view focusing on the O – Si – O – Al – Al – Al atomic layers around the interface. Large grey balls indicate interfacial Al layer atoms. Shaded large grey balls belong to deeper Al layers.

The supercell calculations were performed using Density Functional Theory calculations as implemented in the Vienna Ab initio Simulation Package (VASP) with the Projector Augmented Wave (PAW) framework[18]. The plane wave cut-off energies have been set to 400 and 606 eV for the smooth and augmentation parts respectively. Reciprocal space integrations are performed on a Gamma centered 5×5×1 k-point mesh. Convergence criteria for the total energies and forces on atoms are set to 0.01 meV and 10 meV/Å respectively. Dipole corrections have been applied to cancel dipole-dipole interaction between periodic images of the supercell[19,20]. The Local Density Approximation has been used for all calculations.

The application of an electric field $\vec{F}$ across a polarizable media can lead to the displacement of charges, electrons and nuclei. Hence, an additional component to the total energy $E_{field} = -\vec{F}.\vec{d}$ is associated to the electric field, where $\vec{d}$ is the slab dipole moment. In order to discuss the effect of external electric fields on the interface alone, we remove this additional contribution from the calculated total energies. By taking the total energy of the slab without the application of an electric field $E_{tot}^0$ as a reference, we define the relaxation energy as

$$E_{relax} = (E_{tot} - E_{field}) - E_{tot}^0 \qquad (1)$$

Equation 1 describes the change in total energy caused by the displacement of charges under an applied electric field. In linear materials, the dipole moment is a linear function of the applied electric field and $d \propto \varepsilon \varepsilon_0 F$, where $\varepsilon_0$ is the vacuum permittivity, and $\varepsilon$ the dielectric constant of the material. In the simple capacitor model, the dielectric constant informs about the capacitance of the material $C \propto \varepsilon \varepsilon_0 / L$, which represents its ability to store charges at its interfaces, separated by a distance $L$.

To study the energetics of atomic bonds at the interface, atomic binding energies are calculated. The binding energy of an interfacial oxygen atom is given by

$$E_{binding}^O = E_{defect} - (E_{ideal} - \mu_O) \qquad (2)$$

where $E_{defect}$ is the total energy of the supercell containing an interfacial oxygen vacancy $V_O$, $E_{ideal}$ is the defect free supercell total energy, and $\mu_O$ is the energy of an isolated oxygen atom. Hence, the binding energy expresses the energy stored in the bonds formed by the oxygen atom, by comparing the bonded ($E_{ideal}$) and dangling bond ($E_{defect}$) cases. The binding energy as defined by Eq. 2 can therefore be compared to the adhesion energy, which relate bonded and unbonded slab energies. This binding energy is a special case of the more general point defect formation energy concept. In a defect formation energy, $\mu_O$ would correspond to the chemical potential of oxygen atoms, that represent the energy of an oxygen atom in a reservoir under experimental temperature and pressure conditions. In comparison, the binding energy only expresses the bond energy between atoms.

To study the details of atomic bonding at the interface, Bader charge analyses have been performed using the Henkelman group code[21]. They inform about charges on atoms, and by extension the charge transfer between atoms at the interface. It is useful to compare the Bader charge analysis to simple electron and bond counting model to understand atomic electron transfers. In bulk aluminum, each atom has three open shell electrons and form twelve bonds, yielding 1/4 electron per bond. Normal to the interface, corresponding to the (111) plane of aluminum, each metal atom has three available bonds, totalizing nine available electrons in total per supercell interface. Figure 1b shows that only 2/3 of the Al atoms bond to oxygen atoms. Therefore, six metal electrons should be involved in interface bonding according to the counting model.

**RESULTS**

In the following, the effect of the morphology of the metal/oxide interface on the mechanical and electronic properties is studied. To extract the effects of different atomic environment at the interface, two cases are considered: (i) an interface characterized by oxygen-metal bonds, and (ii) an interface characterized by less oxygen-metal bonds and more cation-metal bonds using interfacial oxygen vacancies. For simplicity, we will refer to these cases as "defect free" and "defective" cases respectively. It has to be noted that these denominations do not refer to the quality of the metal/oxide interface here. Indeed, in the context of an amorphous oxide, both oxygen-metal and cation-metal bonds will be present at the interface. We simply use the oxygen-terminated oxide (Fig. 1a) as the reference. The link between bond formation, atomic charge transfer, and interface adhesion is evidenced. The effect of an electric field $\vec{F}$ applied normal to the interface plane (as shown in Figure 1a) on the electronic properties of interface as a function of morphology is also explored.

1) Interface charge transfers and bonding configuration

The charge transfers across the metal/oxide interface have been determined using Bader charge analysis. Figure 2a shows the net Bader charge in the oxide part of the slab as a function of the applied electric field. At zero electric field, the net charge in the oxide is negative and the metal transfers about 6 electrons to the oxide. It is interesting to note that a simple electron and bond counting model also concludes that the aluminum slab would give 6 electrons to interface bonding per supercell. This result is very close to the calculated net Bader charge in oxide and suggests that aluminum loses its interface bonding electrons in favor of $SiO_2$ via the interfacial oxygen atoms. This is compatible with the fact that Al has the smallest electronegativity (1.61) among Si (1.90) and O (3.44) atoms. On the other hand, Figure 2b shows that the charge on oxygen atoms at the interface remains quasi bulk-like, somehow saturated. Oxygen atoms have the strongest electronegativity in the supercell and are likely to attract bonding electrons. Thus, Al atoms cannot transfer electrons to the saturated oxygen atoms but can transfer them to the second nearest neighbors, corresponding to an Si atom layer. This phenomenon also occurs in the oxygen vacancy case, suggesting the existence of interfacial second nearest neighbor Si-Al bonds.

In presence of an applied electric field, the net charge in oxide decreases. This implies that the electric field enhances the charge transfer from the metal to the oxide. This transfer occurs at a rate of 0.25 and 0.27 electron/V.Å$^{-1}$ per supercell in the defect free and defective interfaces respectively. For the defective case, the oxide layer near the interface is less charged than that for the defect free case, illustrating a lesser charge transfer of about 0.8 electron per supercell. As seen in Figure 2b, the charge redistribution at the interface is very localized, limiting the transfer to the first two atomic layers. This indicates that the interfacial charge transfer and its difference between the defect free and defective cases is controlled by the interface bonding structure only.

In bulk SiO$_2$, Si atoms are tetrahedrally coordinated and loose ~0.78 electron per bond in favor of oxygen atoms. Using these numbers as reference, the inter atomic electron transfers at the interface are deducted from the net Bader charges, and summarized in Figure 2c. The net Bader charge of Si and Al atoms facing the oxygen vacancy suggest an electron transfer from the metal layer directly to the Si atoms. The 0.8 electron difference in charge transfer observed between the defect free and defective cases (Figure 2a) is roughly the amount of electron that is not collected by the missing oxygen atom from the metal layer. Consequently, the charge transfers of Figure 2c are compatible with a bonding model where interfacial Si atoms restore their tetrahedral-like coordination by bonding with Al atoms directly, in both the defect free and the defective cases. This interfacial Si-Al bond is not predicted by the simple electron and bond counting model that only account for first neighbors. Then, the agreement on the transfer of about 6 electrons across the interface between the simple model and the Bader charge analysis is actually fortuitous. More details on bonding and interface states are available in the Supplementary Material.

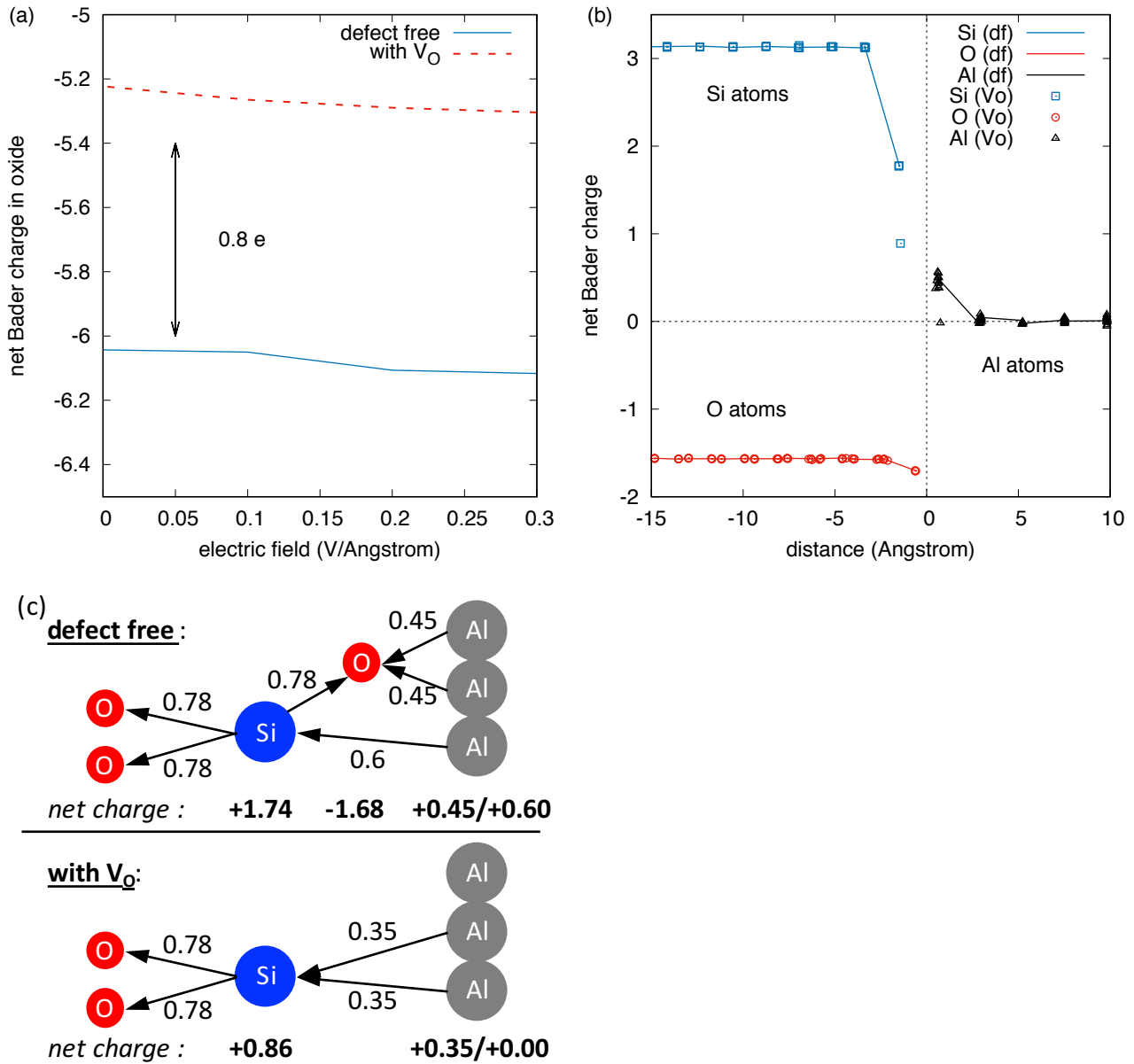

Figure 2: (a) Net charge of the oxide part of the supercell as a function of the applied electric field for the defect free (solid line) and defective (dashed line) cases. (b) Net charge on atoms as a function of their distance to the interface without external electric fields, for the defect free (solid line) and defective (symbols) cases. The vertical dotted line indicates the position of the interface. (c) Electron transfer at the Al/SiO$_2$ interface and net charge of interfacial atoms.

2) Interface response to an electric field

The variation of the net Bader charge in the oxide as a function of the applied electric field shows the response of the metal/oxide interface. The interface exhibits a built-in dipole moment even in the absence of external electric field, illustrating the charge transfers associated with the alignment of Fermi levels during the formation of the interface. The calculated built-in dipole moments are 0.39 and 0.36 e.Å for the defect free and defective cases respectively. This result is consistent with the lesser charge transfer associated with the presence of oxygen vacancies at the interface (Figure 2a). The supercell dielectric constants are 7.16 and 7.29 for the defect free and defective cases respectively. The larger value for the defective case suggests that the interfaces with oxygen vacancies exhibit larger capacitance. This result is misleading since no additional localized interface state is created with interfacial oxygen vacancies (see Supplementary Material). Therefore, the oxygen vacancies do not create additional storage space for electrons at the interface, and do not create dangling bonds. The larger dielectric constant emerges from the altered bonding character at the interface, having less Al-O polar bonds and more "metallic-like" Al-Si bonds. The overall system being subtly more metallic-like, the supercell dielectric constant increases. This is illustrated by the calculation of Born effective charges of Si atoms, that yields a 40% decrease going from the bulk-like part of the oxide to the interface. The response of Si atoms to an electric field is therefore decreased at the interface, evidencing the more metallic-like behavior of the interfacial oxide atomic layers. In comparison, Si-O bonds in bulk are more polar, and exhibit a stronger response to electric fields.

3) Interface energetics

The effect of an applied electric field has also been characterized by the relaxation energy (Equation 1) shown in Figure 3. The negative relaxation energy means that the electric field stabilizes the interface. This is consistent with our previous results, where the electron transfer from the metal to the interfacial Si atom forms a bond, that helps the Si atom to keep its preferred tetrahedral coordination. This charge transfer is further aided by the application of the electric field that further decreases the relaxation energy. The slope of the relaxation energy as a function of the applied electric field is steeper in the defective case, which contains an additional interfacial Al-Si bond per supercell. Given the very limited amount of interface reconstruction around the oxygen vacancy (See Supplementary Material), this result suggests that the transfer of metal electrons to Si atoms plays a role in stabilizing the metal/oxide interface. This role is small, when comparing the electric field relaxation energy to the binding energy (see Eq. 2) of an oxygen atom at the interface (8.54 eV).

The calculated interface and adhesion energies are listed in Table I. The defective case exhibits a smaller interface energy than the defect free one. Since there are less Al-O bonds per unit area in the defective case, this shows that Al-O bonds are more expensive to create than Si-Al bonds at the interface. The ratio between the adhesion energies of the defective and defect free case is roughly 3/4, which is the ratio of the number of interfacial oxygen atoms per unit area. Therefore, the adhesion energy is roughly proportional to the number of interfacial Al-O bonds, which further illustrates the very limited amount of interface reconstruction around interfacial oxygen vacancies. Consequently, Al-O bonds are expected to play a dominant role in the interface bonding.

The interface of the supercell depicted in Figure 1b can be divided into four sub-units. The defect free sub-units have 1 Si-Al and 2 O-Al bonds, while sub-units with an oxygen vacancy have 2 Si-Al bonds. A side view of these sub-units is shown on Fig. 2c. Assuming that all Si-Al bonds are equivalent, and comparing the adhesion energy for the defect free and defective cases, the calculation of the Si-Al and O-Al interfacial bond energy gives 0.17 eV and 1.32 eV respectively. The interfacial Si atom around the oxygen vacancy only weakly contributes to interface adhesion. In bulk $SiO_2$, the calculated binding energy of the oxygen atom is 9.08 eV, yielding 4.54 eV per Si-O bond. It is interesting to note that the binding energy of an interfacial oxygen atom (8.54 eV) deviates from the simple sum of bond energies ($E_{bond}^{Si-O} + (2 E_{bond}^{O-Al} + E_{bond}^{Si-Al}) - 2E_{bond}^{Si-Al} = 7.24$ eV). This illustrates the importance of the interfacial charge transfers, responsible for the non bulk-like nature of the Si-O bonds at the interface. The calculated oxygen atom binding energies and energy of bonds are summarized in Table II.

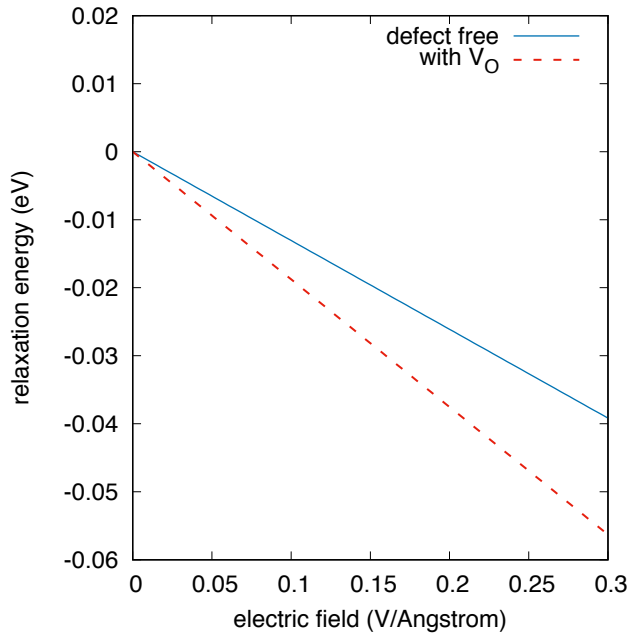

Figure 3: Relaxation energy (defined in Eq. 1) as a function of the applied electric field, for the defect free (solid line) and oxygen vacancy cases (dashed line).

|  | $E_{int}$ (meV/Å$^2$) | $E_{adh}$ (meV/Å$^2$) | dE/dF (meV.Å$^{-2}$/V.Å$^{-1}$) |
|---|---|---|---|
| defect free | 80.3 | -134.6 | -1.561 |
| with V$_O$ | 66.3 | -105.0 | -2.243 |

Table I: SiO$_2$/Al interface energy E$_{int}$, adhesion energy E$_{adh}$, and variation with electric field dE/dF in the defect free and oxygen vacancy cases.

| | $E^O_{binding}$ | $E^{Si-O}_{bond}$ | $E^{Si-Al}_{bond}$ | $E^{O-Al}_{bond}$ |
|---|---|---|---|---|
| SiO$_2$ bulk | 9.08 | 4.54 | - | - |
| SiO$_2$/Al interface | 8.54 | - | 0.17 | 1.32 |

Table II: Binding energy of oxygen atoms $E^O_{binding}$ (Eq. 2) and energy of bonds $E_{bond}$ in bulk SiO$_2$ and at the SiO$_2$/Al interface. Energies are given in eV, for the F=0 V/Å case.

4) Interface morphology and electrical properties

The interface exhibit both polar Al-O and "metallic-like" Al-Si bonds, with the metal electrons spilling to the oxide side. Less electron transfer from the metal to the oxide occurs in the presence of interfacial oxygen vacancies (Figure 2), where the lesser amount has been linked to the absence of strongly electronegative oxygen atoms. Figure 4 displays the planar averaged total potential in the vicinity of the interface, where the interface position has been deducted from atomic positions. An increase in the potential around the vacancy position (dashed line) is evidenced. The metal potential invades the oxygen vacancy region, which virtually decrease the oxide thickness by one atomic layer. The potential and charge density are linked together via the Poisson equation. The aforementioned "metal potential invasion" reveals that electrons at interfacial oxygen vacancy sites feel a metallic environment. This effect is more pronounced in the extreme case of maximal interfacial oxygen vacancy concentration, which corresponds to a Si-terminated interface, containing only Si-Al bonds. In this case, the atomic relaxation is also very limited and the mismatch between the interface location defined by atomic positions and defined by the total potential is even more clearly illustrated.

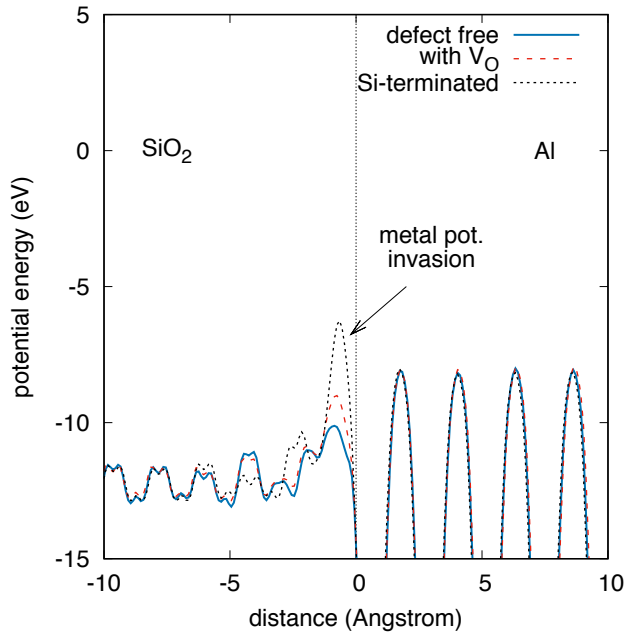

Figure 4: Planar averaged total potential as a function of the distance to the interface for the defect free case (solid line), oxygen vacancy case (dashed line), and Si-terminated interface (dotted line). The bump illustrating the metal potential invasion is located close to the interfacial oxygen layer position. The interface defined by atomic positions is set at 0 distance, located by the vertical dotted line.

**DISCUSSION**

The Al/SiO$_2$ system has been studied as a typical metal/oxide interface. The Bader charge analysis at the interface revealed that a charge transfer occurs from the metal to the oxide. This charge transfer extends to the second atomic layer in the oxide leading to a modification of the oxide properties at the interface such as Si-O bonding. This is illustrated by the difference between the calculated binding energies of oxygen atoms at the interface, and the one predicted from atom bond energies using the bulk value for the Si-O bond. Oxygen binding energies (Eq. 2) and defect formation energies are defined up to a constant, which is the oxygen chemical potential $\mu_O$. Since this potential is not a variable in our calculations and acts simply as a constant offset, the difference between the interfacial and bulk-like oxygen binding energies, equals the formation energy difference. Therefore, oxygen vacancies exhibit a formation energy at the metal/oxide interface that is 0.55 eV lower compared to the vacancy formation in bulk. Consequently, the interface constitutes a weak point where oxygen vacancies are more likely to form. This result is supported by Ref. 22 where the entropy contribution to the defect formation Gibbs free energy is accounted for. The presence of interfacial oxygen vacancies also lowers the interface energy. This reveals that cation-metal bonds are favored over oxygen-metal bonds at the interface.

The presence of Si-Al bonds confers to the interface a metallic character. This metallic character is further amplified by the presence of interfacial oxygen vacancies that doubles the number of Si-Al bonds. This metallic character is illustrated by the metal potential invasion in the first oxide atomic layer, and the larger supercell dielectric constant. It is worth noting that no atomic diffusion occurs in the present case, differing from the metal atom diffusion observed at Cu/low-k dielectric interfaces[23]. Hence, the observed metal potential invasion is a purely electrostatic effect, mediated by the interfacial Si-Al bonds. The metal potential invasion leads to a mismatch between the atomic definition of the interface position, and the electrostatic definition. For the electrons, that feel the electrostatic effects, interfacial cation-metal bonds lead to a virtual thinning of the oxide.

In the percolation path scenario of dielectric breakdown, a collection of point defects creates a conducting path across the dielectric allowing electrical current to flow. This model properly reproduced the lowered critical defect density observed at breakdown when the oxide thickness is decreased, or when the voltage is increased across ultrathin oxides[3]. However, the lower defect density observed at breakdown in thin oxides can also be caused by a higher density of interfacial cation-metal bonds. Indeed, the metallic potential invades the oxide via the interfacial Si-Al bonds, effectively thinning the oxide. This situation is similar to the invasion percolation mediated by the diffusion of metal atoms through low-k dielectrics[23], but without the atomic diffusion in the present case. For ultrathin oxides (~1 nm), this effective thinning can constitute as much as 10% of the oxide thickness. The interface morphology can then help the formation of a percolation path, without physically changing the oxide thickness. This stresses the importance of the interface morphology in the development of breakdown models for thin oxides.

The virtual oxide thinning can also locally lead to an increase of the oxide electric field. This makes interfacial cation-metal bonds preferential nucleation points for the conductive filament in resistive switching devices. The morphology of the conductive filaments having a strong impact on the set and reset voltages[6-8,16], the careful control of interface morphologies can in turn lead to a better control of ReRAM device performance and stability. Moreover, stable device results can be reproduced by the Quantum Point Contact model if the constriction point exhibits a well behaved potential barrier[4,24]. This implies that the constriction point must have a precise and controlled atomic morphology, stressing the importance of the local atomic environment on the formation of conductive filaments.

The effect of local environment on point defect generation will also be important in amorphous oxides. Due to their atomic structure, metal/amorphous-oxide interfaces exhibit both metal-oxygen and metal-cation interfacial bonds. The ensuing virtual oxide thinning will be even more pronounced in suboxide films, such as used for ReRAM applications. In these materials, the formation of oxygen vacancy clusters is crucial, and primarily nucleate at interfaces or on extended defects such as grain boundaries[5]. The accurate modeling of oxygen atoms diffusion, creating and annihilating oxygen vacancies, requires to account for their local environment (e.g. interface or grain boundary). Failure to do so can introduce a bias in the creation/annihilation dynamics that can be detrimental to the understanding of device stability across Set/Reset cycles. Moreover, an asymmetry in the distribution of interfacial oxygen vacancies between both side of the oxide, and consequently in the density of interfacial cation-metal bonds, could be responsible for the polarity dependent breakdown observed in Metal-Oxide-Semiconductor Field Effect Transistors[25]. An asymmetric metal potential invasion could alter built-in interface dipoles and affect oxide electric fields.

The effect of oxide electric field on point defect generation in $SiO_2$ has been studied in Ref. 2, 13 and 25. In bulk $SiO_2$, oxygen vacancies give birth to Si-Si bonds that create a defect state in the band gap. The strong trapping character of this state is caused by the large relaxation associated with level charging and discharging[26]. At the Al/$SiO_2$ interface, no defect state localized in the band gap could be identified and assigned to interfacial oxygen vacancies. Moreover, no interfacial Si dangling bonds are found. The very little atomic relaxation associated with the creation of interfacial oxygen vacancy supports the non-trapping character of the Si-Al bonds. Moreover, the Fermi level of the system, that corresponds to the Fermi level in the metal, is located inside the oxide band gap. Any conduction electron located at the interface and above the Fermi level could easily flee to the metal. At the interface, the examination of atomic electronegativities, predicts that Si-Al bonds are much less polar than Al-O bonds. The Al-O bond is also more polar than the Si-O bond. Then, according to the field accelerated defect creation model, the Al-O bond is supposed to be more sensitive to electric fields[2,11-13]. However, the interfacial oxygen atom binding energy is reduced by only ~30 meV, by applying a 0.3 V/Å electric field, which corresponds to a ~6.4 MV/cm oxide field. For the same oxide field range, the bond breakage energy in bulk $SiO_2$ decreases by ~0.5eV[13]. This discrepancy stems from the different local environments that modifies the atomic response to electric fields. The interfacial Si and O atoms are in intimate contact with the electrode applying the electrical bias, and have different Bader and Born effective charges than in bulk. Consequently, they do not have the same response to electric fields. Bond breaking models devised for bulk materials cannot be used to describe defect creation close to metal/oxide interfaces because they do not account for the atomic environment difference. This illustrates the need for more elaborate point defect generation models in the interface regions, especially for ultra thin oxides.

**CONCLUSION**

The prototypical Al/SiO$_2$ metal/oxide interface has been studied and its properties explored. The bonding structure has been found to largely dictate the interface electronic properties, where the oxide cation atoms acquire a metallic character and act as electron hosts. This stabilizes the interface via the formation of "metallic-like" cation-metal bonds. Interfacial cation-metal bonds allow the invasion of the metallic potential inside the oxide, virtually decreasing the oxide thickness. This locally enhances the oxide electric fields, generating hot spots for further defect creation. Such interface morphology dependent effects are particularly important in ultra thin and amorphous oxides.

## SUPPLEMENTARY MATERIAL

See supplementary material for deeper study of the interfacial oxygen-metal and cation-metal bonds, as well as the interface electronic structure.

## ACKNOWLEDGMENTS

This work was funded by the Air Force with program name: Aerospace Materials for Extreme Environment, and grant number: FA9550-14-1-0157. We acknowledge Advanced Research Computing at Virginia Tech for providing the necessary computational resources and technical support that have contributed to this work (http://www.arc.vt.edu).

**SUPPLMENTARY MATERIAL**

**Bonding at Al/SiO$_2$ interface**

The Bader charge analysis suggests that an electron transfer occurs between interfacial Si and Al atoms in order for the Si atoms to keep their preferential tetragonal coordination without dangling bonds. To reveal interfacial bonds, the charge density and Electron Localization Function (ELF)[1] have been studied. Figure S1a shows an isosurface of the charge density at the interface. For the sake of clarity, only the charge density inside the Bader volumes associated to interfacial Al atoms is displayed. A flat interface between Bader volumes on the Al-Si axis would indicate a frontier between both atom electron clouds, revealing their overlap. Therefore, a metallic or covalent-like Al-Si bond would be indicated by a charge density lobe emerging from Al atoms truncated by a plane normal to Al-Si axes. In Figure S1a, two Al-Si bonds around the oxygen vacancy, and only one without the oxygen vacancy, are evidenced. The interfacial Si and Al atom electron clouds stay connected at higher charge density isosurface levels where Al-Al clouds are not.

Figure S1b displays an isosurface ($\eta$=0.5) of the ELF around the interface. Without an oxygen vacancy at the interface, the four lobe pattern is reminiscent of the preferred Si tetrahedral coordination. Three lobes are centered around oxygen atoms, depicting their ability to localize electrons due to their high electronegativity. The fourth lobe reveals an attractor between the interfacial Si and Al atoms. By decreasing the ELF isosurface level from $\eta$=1.0 to 0.6, the fourth lobe merges with the network of channels connecting the Al atoms. The Al-Si lobe position, and its connection to the Al network, suggest that Al-Si bonds are between covalent and metallic types. The same trends are observed in the defective case, where the basin of the Al and Si-Al attractors are largely merged at $\eta$=0.5, which is the signature of electron delocalization, while Si and O ELF lobes are still disconnected. The vacancy region appears like a bump in the Al sea. At the same isosurface level, Al and O bassins are still not merged, illustrating the polar nature of the Al-O bond.

The charge density and ELF analysis are compatible with the charge transfers observed in the previous section. The observed Si-Al bond lengths are 2.44 Å in the defect free case, and 2.47 Å and 2.57 Å in the oxygen vacancy case. The amount of interface reconstruction associated with the oxygen vacancy is very limited, with the neighboring Si atom approaching the Al layer by only 0.09 Å. Hence, the two Si-Al bonds of the vacancy case exhibit slightly different lengths. These bond lengths are comparable with the covalent bonds observed in bulk silicon (2.35 Å) and metallic bonds in bulk aluminum (2.86 Å).

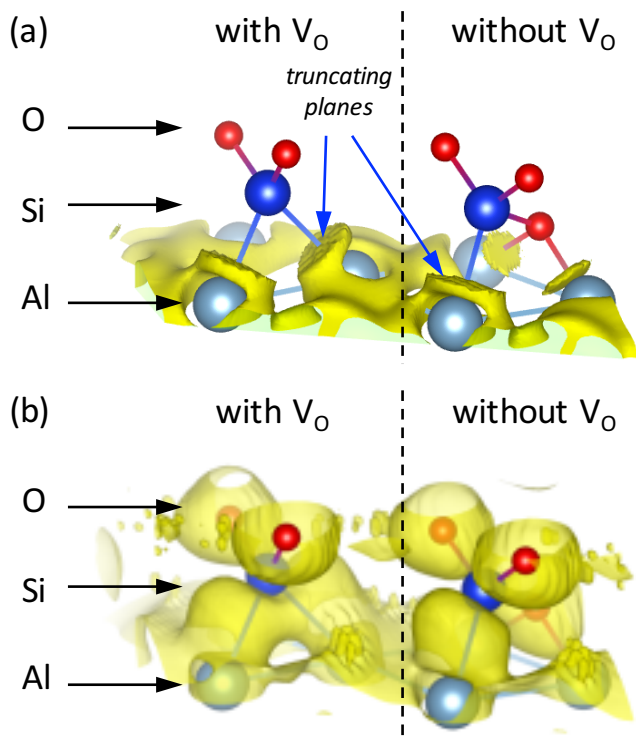

Figure S1: (a) Isosurface of the charge density belonging only to the interfacial Al layer as determined by the Bader charge analysis. (b) Isosurface of the Electron Localization Function ($\eta=0.5$) at the SiO$_2$/Al interface.

**Electronic states**

At metal/oxide interfaces, the tunneling of metal electrons inside the dielectric band gap forms Metal Induced Gap States model (MIGS). The associated local charge transfer from the metal to the dielectric smoothly align the Fermi level of both constituent at the interface[2]. This concept can be visualized by analyzing the wave function projection of the supercell electronic states onto each atom PAW sphere. To this end, the wave function s, p and d character of bulk SiO$_2$, and bulk Al, electronic states are computed separately using unit cells and very dense k-point meshes. This provides a database of s, p and d signatures of bulk states. For each supercell electronic state, the supercell wave function character is tested against bulk SiO$_2$, and bulk Al database. The signatures comparison is carried out combining both the amplitude of s, p and d projection, and their relative proportion in a maximum likelihood criterion. The use of bulk references enables the determination of bulk-like SiO$_2$ and Al states, and identity interface states by elimination. This is particularly useful in supercell calculations, where band folding generates a large number of electronic states at any given k-point of the supercell Brillouin Zone, blurring out band structure features that could be used to identify the states. The real space information (atom position), usually lost in conventional density of states studies, has been used to determine the spatial extent of bulk-like and interface states. Electronic state boundaries are defined so that 90% of the wavefunction is contained inside the boundaries. This criterion helps eliminating long vanishing wavefunction tails and inform about the state localization, and hence their nature.

The study of extended structures using supercells generates a large amount of electronic states, at a given k-point in the supercell Brillouin Zone, due to band folding. A 90% likelihood criterion has been used to identify bulk-like states, as shown in Figure S2a. Figure S2b exhibits the remaining states, those that are not categorized as bulk-like SiO$_2$ or bulk-like Al. In the following, we will refer to them as mixed states, since their interface nature has not been established at this point. The Fermi level is set as the origin of the energy axis, and corresponds to a metal state. Deep oxide valence states below -20 eV are well localized inside the oxide part. However, oxide valence states closer to the $SiO_2$ Valence Band Maximum (-3.7 eV) tend to spill inside the metal part. Some metal states also spill into the oxide (MIGS), but to a lesser extent. Most of the mixed states exhibit a large degree of delocalization normal to the interface. Mixed states only exist at energies where Al and $SiO_2$ valence bands overlap, which is a necessary condition for state coupling. In $SiO_2$ band gaps, mixed states originating from the metal side penetrate through 2-3 atomic layers inside the oxide. They also exhibit a large degree of delocalization in the metal. Only a few mixed states in the -7/-8 eV energy window, in a $SiO_2$ valence band gap, are localized at the interface and could truly be identified as interface states. The $SiO_2$ Conduction Band Minimum (2.3 eV) along with most of the oxide conduction bands exhibit a strong coupling to metal states and cannot be characterized as bulk-like. Indeed, an oxide conduction electron above the Fermi level is not confined in $SiO_2$ due to the absence of potential barrier.

Under the application of an electric field (F=0.3 V/Å in Figure S2c and S2d), the supercell energy spectrum overcomes some changes. Band bending can be seen on the bulk-like $SiO_2$ valence bands, where the metal corresponds to higher potential energy. The most noticeable change affects the mixed states close to the Fermi level, and close to the $SiO_2$ Valence Band Maximum, where metallic mixed states penetrate deeper inside the oxide. Using the same 90% likelihood criterion, less bulk states are mistaken for mixed states. This shows a stronger metal/oxide states coupling that effectively separate the bulk states from the mixed ones. This is seen on Figure S2d where mixed states below the Fermi level tend to be more centered around the interface. This comes in favor of the metallic-like interface bonding model, where Al electrons penetrate in the oxide through the Si atoms.

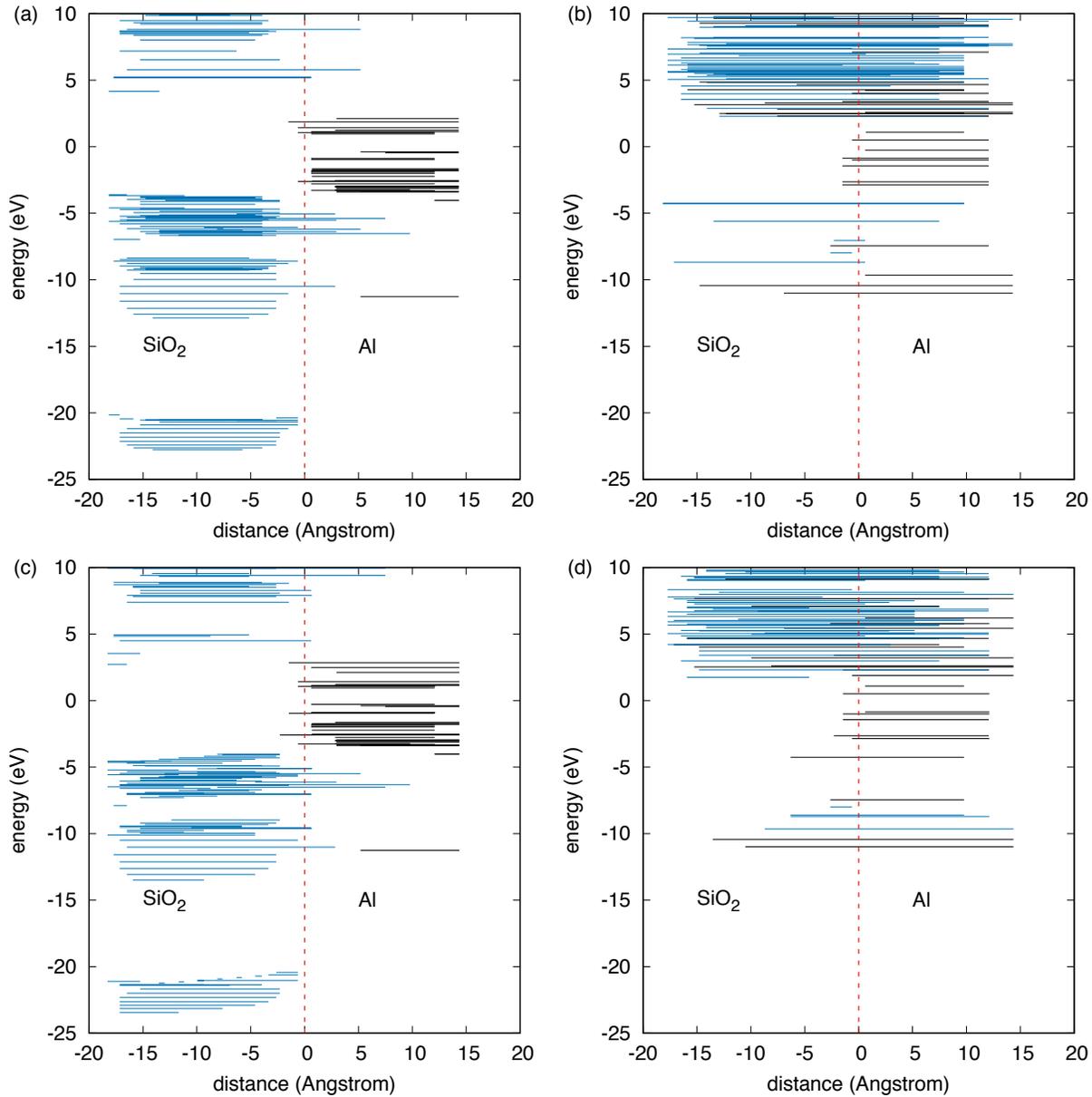

Figure S2: Decomposed electronic structure at the Γ point for F=0 V/Å (a,b) and F=0.3 V/Å (c,d), showing bulk-like states (a,c), and mixed states (b,d). For the sake of clarity, a supercell with a 20.6 Å$^2$ defect free interface has been used in order to limit the amount of band folding. The Fermi level corresponds to zero energy. The vertical dashed line indicates the position of the interface.

A detailed analysis of the charge density has been performed to assess the possible localization of electrons at the Si-Al interface bond. The charge density of all electronic state has been integrated separately, accounting only for the region between the interfacial Al and Si atoms. The result, the amount of electron on the Al-Si bond for each electronic state is displayed on Fig. S3. A few states exhibit a high amount of electrons on the bond, indicated by circles. However, for each of these state, the visualization of the full charge density map does not reveal localization at the interface. States with energies below the SiO$_2$ VBM are delocalized in the oxide. States with energies above the SiO$_2$ VBM are delocalized in the metal. This illustrates the non-trapping character of the interfacial Si-Al bond.

In the case of oxygen vacancies at the interface, the same behavior is observed. Interfacial oxygen vacancies do not create trapping states.

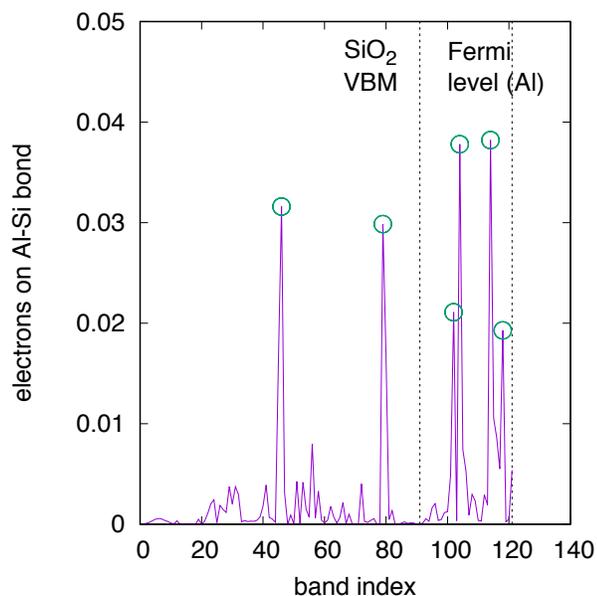

Figure S3: Amount of electrons located between the interfacial Si and Al atoms (solid line). Circles indicate the states with the highest electron concentration between the interfacial Si and Al atoms.

**References**

[1] A. Savin, R. Nesper, S. Wengert, T. F. Fässler, *Angew. Chem. Int. Ed. Engl.* **36**, 1808 (1997)
[2] W. Monch, *J. Vac. Sci. Technol. B* **17**, 1867 (1999)